\documentclass{PoS}
\rightline{\parbox{4cm}{\large\rm
Edinburgh 2006/35\\
SWAT 05/476\\
SHEP-0634
}}

\usepackage{amssymb}
\usepackage{lscape}
\usepackage{longtable}
\usepackage{epsfig}

\usepackage{graphicx}
\usepackage[figuresright]{rotating}

\newcommand{\AmS}{{\protect\the\textfont2
  A\kern-.1667em\lower.5ex\hbox{M}\kern-.125emS}}

\title{Light meson masses and non-perturbative renormalisation in 2+1 flavour domain wall QCD}

\ShortTitle{Light meson masses and non-perturbative renormalisation in 2+1 flavour domain wall QCD}

\author{RBC and UKQCD Collaborations: C.~Allton$^e$, D.J.~Antonio$^a$,
P.A.~Boyle$^a$, C.~Dawson$^b$, A.~J\"uttner$^d$, R.D.~Kenway$^a$, S.~Li$^c$,
M.~Lin$^c$, C.M.~Maynard$^g$, J.~Noaki$^d$, B.J.~Pendleton$^a$,
\speaker{R.J.~Tweedie$^a$}\thanks{(E-mail: \tt{rjt@ph.ed.ac.uk}).}, ~A.~Yamaguchi$^f$ and
J.~Zanotti$^a$.\\ 
\llap{$^a$}SUPA, School of Physics, The University of Edinburgh,
Edinburgh EH9 3JZ, UK\\
\llap{$^b$}RIKEN BNL Research Center, Brookhaven National Laboratory,
Upton, NY 11973, USA\\
\llap{$^c$}Physics Department, Columbia University, New York, NY
10027, USA\\ 
\llap{$^d$}School of Physics and Astronomy, The University of
Southampton, Southampton SO17 1BJ, UK\\
\llap{$^e$}Department of Physics, University of Wales Swansea, Swansea
SA2 8PP, UK\\
\llap{$^f$}Department of Physics and Astronomy, The University of
Glasgow, Glasgow G12 8QQ, UK\\
\llap{$^g$}EPCC, School of Physics, The University of Edinburgh,
Edinburgh EH9 3JZ, UK}

\abstract{ We present results for the light meson masses, the bare
strange quark mass and preliminary non-perturbative renormalisation of
$B_K$ in 2+1 flavour domain wall QCD. The ensembles used were
generated with the Iwasaki gauge action and have a volume of
$16^3\times32$ with a fifth dimension size of 16 and an inverse
lattice spacing of 1.6 GeV. These ensembles have $u$ and $d$ masses as
low as one quarter of the strange quark mass. All data were generated
jointly by the UKQCD and RBC collaborations on QCDOC machines. }

\FullConference{XXIVth International Symposium on Lattice Field Theory\\
                July 23-28, 2006\\
                Tucson, Arizona, USA}    

\begin{document}

\section{INTRODUCTION AND SIMULATION PARAMETERS}

This work describes results from 2+1 flavour domain wall
ensembles with a lattice size of $16^3 \times 32$ recently generated
by the RBC and UKQCD collaborations. The analysis was carried out
using the standard domain wall Dirac operator~\cite{Furman:1994ky} and
Pauli-Villars fields with the action introduced
in~\cite{Vranas:1997da}. The configurations were generated with the
Iwasaki gauge action~\cite{Iwasaki:1983ck}, using the
RHMC algorithm~\cite{Clark:2004cp,bob} with a trajectory length of 1.0,
domain wall height of 1.8, $\beta = 2.13$ and a fifth dimension of
length 16. Three ensembles were generated with a fixed approximate
strange quark mass, $am_s = 0.04$, and a light isodoublet with masses
$am_{ud}$ = 0.01, 0.02 or 0.03. The number of trajectories in each
ensemble and the number of configurations used in the analysis are
shown in table~\ref{tab:datasets}. At present, $24^3\times 64$
ensembles with the same parameter values are being generated in order
to study finite size effects and to compute a wider range of physical
quantities.

In this paper we consider meson correlators with valence quark masses
equal to the light isodoublet in the sea, the strange quark mass in
the sea and the non-degenerate combinations. To improve our
statistics, correlators were oversampled and averaged into bins of
size between five and ten, depending on the Monte Carlo time
separation between measurements. The binning is consistent with the
integrated auto-correlation length for the pseudoscalar meson, which
was calculated to be of order 50 trajectories. Multiple sources per
configuration and several different types of smearing have also been
used to improve the signal. A full correlated analysis was then
performed with the binned data as input.

\section{RESULTS}

\subsection{The residual mass}

The residual mass, which is a measure of the violation of chiral
symmetry due to the finite fifth dimension~\cite{Furman:1994ky}, was
calculated from the ratio of the point-split pseudoscalar density,
$J_{5}$, at the middle of the fifth dimension, to the pseudoscalar
density, $P$, built from the fields on the walls~\cite{Furman:1994ky}
\begin{equation}
am_{\rm res} = \frac{ \Sigma_{\vec{x},\vec{y}} \langle
J_5(\vec{y},t)P(\vec{x},0)\rangle }{\Sigma_{\vec{x},\vec{y}} \langle
P(\vec{y},t) P(\vec{x},0)\rangle}.
\end{equation}
A simple unitary linear chiral extrapolation was performed, $am_{\rm
res} = A + B am_f$, where $am_{f}$ is the input valence quark mass and
the self-consistent value, $am_{\rm res}^{\chi} = \frac{A}{(1+B)}$, was
defined to be the residual mass in the chiral limit. The
self-consistent value obtained in the chiral limit is given in
table~\ref{tab:datasets}. This corresponds to a residual mass of 5
MeV.

\begin{table}[!b]
\begin{center}
\begin{tabular}{cccccccc}
\hline
$V\times L_s$ & $\frac{am_{ud}}{am_{s}}$ & $N_{\rm traj}$ & $N_{\rm meas}$ & $N_{\rm meas}^{\rm NPR}$ & $am_{\rm res}^{\chi}$ & $am_s$ & $am_{PS}$\\
\hline
$16^3 \times 32 \times 16$ & 0.01/0.04 & 4000 & 1400 & 150 & & & 0.245(2)\\
$16^3 \times 32 \times 16$ & 0.02/0.04 & 4000 & 700 & 150 & 0.00308(3)& 0.042(2) & 0.324(2)\\
$16^3 \times 32 \times 16$ & 0.03/0.04 & 4000 & 700 & 150 & & & 0.390(2)\\
\hline
\end{tabular}
\end{center}
\caption{\label{tab:datasets} RHMC 2+1 flavour datasets, the value of
the residual mass in the chiral limit, $am_{\rm res}^{\chi}$, the bare
strange quark mass and the unitary pseudoscalar meson mass, $am_{PS}$.}
\end{table}

\begin{figure}
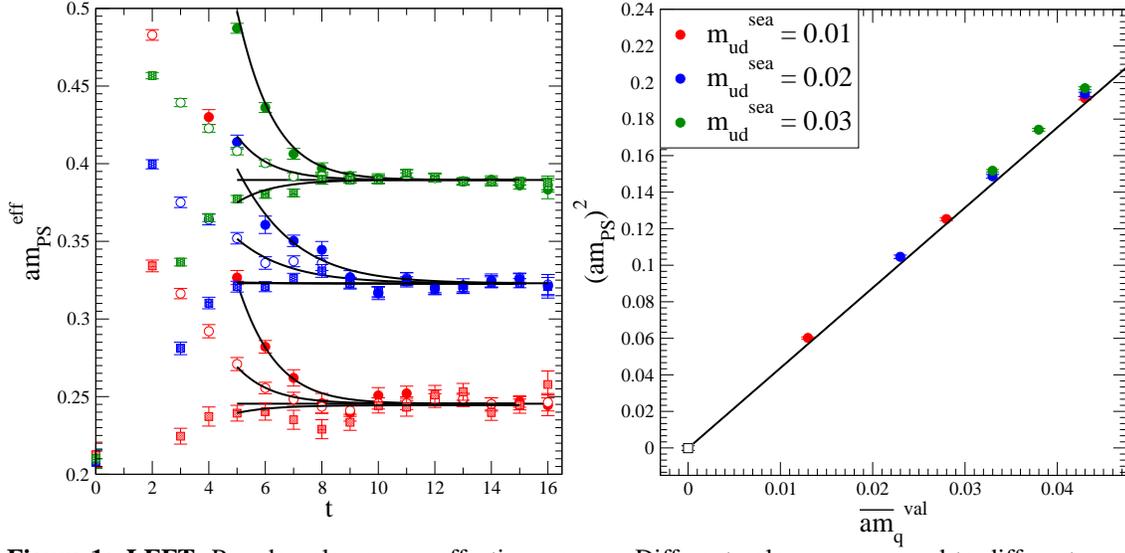

{\raisebox{+0.4cm}{\includegraphics[width=.49\textwidth]{PSeffm2.eps}}}
\includegraphics[width=.49\textwidth]{strange.eps}
\vspace*{-0.5cm}
\caption{\label{plot:ps}{\bf LEFT:} Pseudoscalar meson effective
  masses. Different colours correspond to different quark masses,
  open, closed and square symbols denote different
  smearings. Horizontal lines show the fit to the ground state mass
  while the curved lines result from inputting the fit parameters into
  equation~(\protect\ref{eq:fitfunc}).
  \label{plot:strange}{\bf RIGHT:} Chiral extrapolation of the pseudoscalar
  meson mass against average valence quark mass $\bar{m_q} = \left(
  am_{q_1}^{\rm{val}} + am_{q_2}^{\rm{val}} \right)/2$. The slope shown
  comes from parameter $B$ in equation~(\protect\ref{eq:psextrap}) and
  corresponds to extrapolation to the chiral limit. Different colours
  denote different light sea quark masses. It may be seen that the sea
  quark mass dependence, parameter $C$ in
  equation~(\protect\ref{eq:psextrap}), is very mild.}
\end{figure}

\subsection{Light meson masses}

Pseudoscalar and vector meson masses were extracted by performing a 
simultaneous double cosh fit to the form
\begin{eqnarray}\label{eq:fitfunc}
C_1(t) &=& A_1^G \left( e^{-m_G t} + e^{-m_G(T-t)}\right) + A_1^E \left( e^{-m_E t} + e^{-m_E(T-t)}\right)\\
C_2(t) &=& A_2^G \left( e^{-m_G t} + e^{-m_G(T-t)}\right) + A_2^E \left( e^{-m_E t} + e^{-m_E(T-t)}\right)\nonumber
\end{eqnarray}
to extract both the ground and first excited states. This helped to
remove the systematic error in the choice of fit range due to the
excited state. Figure~\ref{plot:ps} (left) shows typical effective
masses for the pseudoscalar mesons.

We set the scale in three ways. Firstly, the lattice spacing was
obtained from $m_{\rho}$ in the chiral limit by performing a simple
linear unitary chiral extrapolation of the vector meson mass
\begin{equation}
am_{V} = C(am_{q}^{val=sea}) + D 
\end{equation}
where the quark mass, $am_q$, was defined by $am_{q} \equiv a(m_{f} +
m_{\rm res}^{\chi})$ with $am_f$ the input quark mass. Secondly, $r_0$
was determined from the quark-antiquark static potential presented in
this conference~\cite{Li:2006gr}, and finally, the method of
planes~\cite{Allton:1996yv} was used. All three methods gave a
consistent value for the lattice spacing of $a^{-1} = 1.60(3)$ GeV,
and therefore a box size of $\sim 2$fm. The lightest pseudoscalar
meson has a mass of approximately $390$ MeV.

The non-degenerate pseudoscalar meson masses were chirally
extrapolated by fitting both the valence and sea quark mass dependence
using the form
\begin{equation}\label{eq:psextrap}
(am_{PS})^2 = A 
	+ B \left( am_{q_1}^{\rm{val}} + am_{q_2}^{\rm{val}} \right) 
	+ C am_{q}^{\rm{sea}}.  
\end{equation}
Figure~\ref{plot:strange}(right) shows the pseudoscalar meson mass
squared plotted against the average valence quark mass,
$\overline{am_q} = (am_{q_1}^{\rm{val}} + am_{q_2}^{\rm{val}} )/
2$. The value of $A$ was found to be consistent with zero in good
agreement with the residual mass behaving as an additive
renormalisation to the input quark mass, $m_f$. The normal quark mass,
$am_n$, was obtained from the physical pion mass by setting $am_n =
am_{q_1}^{\rm{val}} = am_{q_2}^{\rm{val}} = am_{q}^{\rm{sea}}$ in
equation~(\ref{eq:psextrap}). Similarly, the strange quark mass,
$am_s$, was obtained from the physical kaon mass by substituting $am_n
= am_{q_1}^{\rm{val}} = am_q^{sea}$ and $am_s =
am_{q_2}^{\rm{val}}$. In both cases the scale was taken to be 1.60(3)
GeV. The value obtained for the bare strange quark mass (see
table~\ref{tab:datasets}) is in agreement with the input value of
$0.04 + am_{\rm{res}}^{\chi}$.

\subsection{Non-perturbative renormalisation for $B_K$}

\begin{figure}
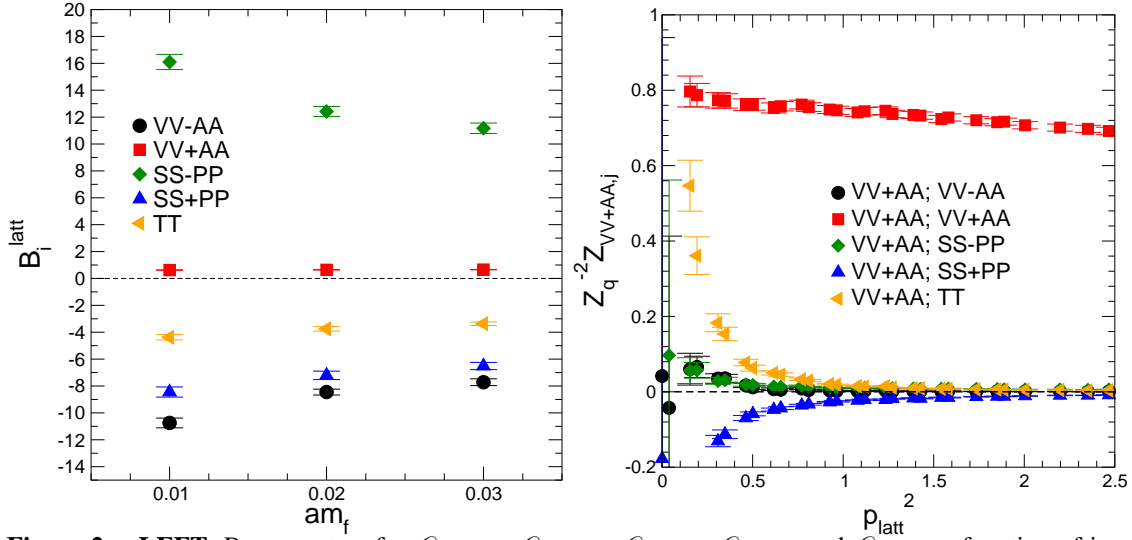

\includegraphics[width=.49\textwidth]{Bpara_all.eps}
\includegraphics[width=.49\textwidth]{ZBpara_all.eps}
\vspace*{-0.5cm}
\caption{\label{plot:Bparam} {\bf LEFT:} B-parameters for
$\mathcal{O}_{VV+AA}$, $\mathcal{O}_{VV-AA}$, $\mathcal{O}_{SS+PP}$,
$\mathcal{O}_{SS-PP}$ and $\mathcal{O}_{TT}$ as a function of input
quark mass, $m_f$. \label{plot:ZB}{\bf RIGHT:} Renormalisation factors
$Z_q^{-2}Z_{VV+AA,j}$ as a function of $p^2_{\rm{latt}}$. This
demonstrates that the mixing matrix is, in practice, block diagonal
with negligible chirality mixing. For each momentum the data has been
linearly extrapolated to the chiral limit ($m_q \rightarrow 0$).}
\end{figure}

$B_K$ measures the QCD correction to the weak mixing between $K^0 =
(\bar{s}d)$ and $\bar{K^0} = (s\bar{d})$ which is an important
constraint on the CKM matrix. It is evaluated from the dimensionless
ratio of the matrix element of the $\Delta S = 2$ operator,
\begin{equation}
\mathcal{O}_{B_K} = \mathcal{O}_{VV+AA} = (\bar{s}\gamma_\mu d )(\bar{s}\gamma_\mu d)
+ (\bar{s}\gamma_5\gamma_\mu d)(\bar{s}\gamma_5\gamma_\mu d) \, ,
\end{equation}
to its vacuum saturation approximation
\begin{equation}
B_K = \frac{\langle\bar{K^0}|\mathcal{O}_{B_K}|K^0\rangle}{\frac{8}{3} m_K^2 f_K^2}.
\end{equation}
Chiral symmetry breaking may cause mixing of the operator
$\mathcal{O}_{B_K}$ with four other four-quark operators
\begin{eqnarray}
{\cal O}_{VV - AA}&=& (\bar{s}\gamma_\mu d)(\bar{s}\gamma_\mu d)
- (\bar{s}\gamma_5\gamma_\mu d)(\bar{s}\gamma_5\gamma_\mu d),\\
{\cal O}_{SS\pm PP}&=& (\bar{s} d)(\bar{s}d)
\pm (\bar{s}\gamma_5 d)(\bar{s}\gamma_5 d),\\
{\cal O}_{TT}&=& (\bar{s}\sigma_{\mu\nu} d)(\bar{s}\sigma_{\mu\nu} d)\,, 
\end{eqnarray}
however, the exponentially accurate chiral symmetry afforded by the
domain wall fermion formulation should strongly suppress this
mixing. In the presence of chiral symmetry breaking the renormalised
$B$-parameter, $B_K^{\rm (ren)}$, is in principle an admixture of the
five lattice $B$-parameters
%
\begin{eqnarray}
B_K^{\rm (ren)} = Z_{VV+AA,VV+AA}/Z_A^2\cdot B_K^{\rm (latt)} 
+ \sum_{i} Z_{VV+AA, i}/Z_A^2\cdot B_i^{\rm (latt)} \,, 
\end{eqnarray}
for $i \in \{VV-AA,SS\pm PP, TT\}$. It may be observed from
figure~\ref{plot:Bparam} (left) that the magnitude of the
$B$-parameters for the wrong chirality operators are large in
comparison with $B^{\rm latt}_{VV+AA}$, the matrix element of
interest, and hence the sizes of the mixing coefficients,
$Z_{VV+AA,i}$, are crucial.

The renormalisation coefficients are evaluated using the
Rome-Southampton method~\cite{Martinelli:1994ty} of non-perturbative
renormalisation. In Landau gauge, the amputated $n$-point correlation
function of the operator of interest, $\Gamma^{(n){\rm latt}}_{\cal
O}$, is constructed by applying, at fixed external quark momenta and
zero quark mass, the condition
\begin{eqnarray}
Z_q^{-n/2}Z_{ij} {\rm P}_k \left[ \Gamma_{{\cal O}_j}^{(n){\rm
latt}} (p\ ;\ a^{-1}) \right] = {\rm P}_k \left[ \Gamma^{(n){\rm
tree}}_{{\cal O}_i} \right] \,,
\end{eqnarray}
where ${\rm P_k}$ represents the application of a particular
spin-color projection and a subsequent trace. 

The set of momenta used to calculate the ratio of renormalisation
factors $Z_{Q^{(\Delta S=2)}}/Z_A^2$ is defined by
\begin{eqnarray}
p_{\rm latt}= \left(\frac{2\pi}{L_x}n_x,\frac{2\pi}{L_y}n_y,
\frac{2\pi}{L_z}n_z,\frac{2\pi}{L_t}n_t\right)
\end{eqnarray}
where $L_x=L_y=L_z=16$ and $L_t=32$, using 1359 combinations of
$(n_x,n_y,n_z,n_t)$ with $-2 \le n_x, n_y, n_z \le 2$ and $-5 \le n_t
\le 5$. These are then averaged into 29 combinations of equal $p_{\rm
latt}^2$. The number of gauge configurations used for the
non-perturbative renormalisation is given in table~\ref{tab:datasets}.

Results for the elements of the top row of the mixing matrix,
$\Lambda_{VV+AA,j}$, $Z_q^{2}Z_{VV+AA, j}^{-1}$, versus $p_{\rm
latt}^2$ are shown in figure~\ref{plot:ZB} (right). It may be seen
from the plot that, in the window of momenta for which contributions
from both hadronic effects (low momenta) and contributions from
discretisation effects (high momenta) are small, the resulting mixing
coefficients are small enough that the matrix becomes block diagonal,
simplifying the inverse, and hence the wrong chirality operators can
be safely neglected. Therefore, we calculate the $B_K$ renormalisation
factor as $Z_{B_K}^{\rm \,RI/MOM}=Z_{VV+AA}/Z_A^2$. In order to obtain
a value for $B_K$ in the $\overline{MS}$ scheme at some scale a
continuum running/matching calculation was performed following the
techniques in~\cite{Aoki:2005ga}. A preliminary value for
$Z_{B_K}(\overline{MS}$,~2GeV$)$ of 0.90(2) was obtained.

\section{SUMMARY}

Recent 2+1 flavour simulations using domain wall fermions, the Iwasaki
gauge action, three different light isodoublet quark masses and a
volume of $16^3 \times 32$ with a fifth dimension size of 16 have been
performed by the RBC and UKQCD collaborations. The chiral symmetry
breaking, parameterised by the residual mass, is consistent with a
small additive quark mass renormalisation of 5~MeV, which is at an
acceptable level with a fifth dimension size of 16. Reasonable signals
have been obtained for the pseudoscalar and vector meson masses. These
ensembles have a lattice spacing of 1.60(3) GeV and hence the lightest
pseudoscalar meson has a mass of approximately 390 MeV. We find that
the bare strange quark mass evaluated from the pseudoscalar chiral
extrapolation is in good agreement with the input strange quark mass.

The improved chiral symmetry afforded by the domain wall fermion
formulation leads to the suppression of mixing with the wrong
chirality operators in the calculation of $B_K$. This greatly
simplifies this calculation and should allow for an accurate
determination of the matrix element~\cite{saulcohen}.

\section*{ACKNOWLEDGEMENTS}
We thank Peter Boyle, Dong Chen, Mike Clark, Norman Christ, Saul
Cohen, Calin Cristian, Zhihua Dong, Alan Gara, Andrew Jackson, Balint
Joo, Chulwoo Jung, Richard Kenway, Changhoan Kim, Ludmila Levkova,
Xiaodong Liao, Guofeng Liu, Robert Mawhinney, Shigemi Ohta, Konstantin
Petrov, Tilo Wettig and Azusa Yamaguchi for developing the QCDOC
machine and its software.  This development and the resulting computer
equipment used in this calculation were funded by the U.S. DOE grant
DE-FG02-92ER40699, PPARC JIF grant PPA/J/S/1998/00756 and by
RIKEN. This work was supported by PPARC grant PP/C504386/1 and PPARC
grant PP/D000238/1. We wish to thank the staff in the Advanced
Computing Facility in the University of Edinburgh for their help and
support for this research programme.

\bibliographystyle{JHEP-2}
\bibliography{refs}

\end{document}